\documentclass[aps,preprint]{revtex4}
\def\beq{\begin{equation}} \def\eeq{\end{equation}}
\usepackage{graphics,epsfig,amssymb}

\begin{document}

\title{Four-nucleon alpha-type correlations and proton-neutron pairing  away of N=Z line}

\author{N. Sandulescu and D. Negrea}
\affiliation{
National Institute of Physics and Nuclear Engineering, P.O. Box MG-6, 76900 Bucharest-Magurele, Romania}
\author{ C. W. Johnson}
\affiliation{Department of Physics, San Diego State University,
 5500 Campanile Drive, San Diego, CA 92182-1233}

\begin{abstract}
We study the competition between alpha-type and conventional pair 
condensation in the ground state of nuclei with neutrons and protons interacting 
via  a charge-independent pairing interaction. The ground state is described by
a product of two condensates, one of alpha-like quartets and the other one of pairs
in excess relative to the isotope with N=Z. It is shown that this ansatz for the
ground state gives very accurate pairing correlation energies for  nuclei with 
the valence nucleons above the closed cores 16O, 40Ca and 100Sn. 
These results indicate that alpha-type correlations are important not only for the 
self-conjugate nuclei but also for nuclei away of N=Z line. In the latter 
case alpha-like quartets coexist with the collective Cooper pairs formed by
the nucleons in excess. 
  
\end{abstract}

\maketitle

%\section{introduction}

It has been suggested long ago that in self-conjugate nuclei proton-neutron pairing 
can induce, through the isospin conservation, four-particle correlations of  
alpha-type \cite{lane}. A related question, repeatedly discussed by various
authors, is whether the ground state of N=Z nuclei can be described as 
a superfluid condensate of alpha-like quartets. 
One of the first models of alpha-type superfluidity in N=Z nuclei was proposed 
by Flowers et al \cite{flowers} and it was based on a BCS-like state made of 
quartets instead of pairs. Recently, this model has been extended by including 
in the BCS state both quartets and pairs \cite{zelevinsky}. 
As any quasi-particle approximations, these models do not conserve exactly the 
particle number. For alpha-type correlations this is a serious drawback 
since in this case the particle number  becomes uncertain in units of 
four particles at a time. Alpha-type condensation in the ground state 
of N=Z nuclei was also studied in particle number conserving 
models \cite{eichler,gambhir,hasegawa,dobes,talmi}. However, majority of these
studies have been done either with schematic single-particle spectra and schematic 
interactions  or using approximations justified for a limited number 
of quartets. A general calculation scheme for taking into account  
alpha-type quartet correlations, valid for any number of quartets and for a general
charge-independent pairing force, was proposed recently in 
Ref. \cite{sandulescu_qcm}. The calculations done in Ref.\cite{sandulescu_qcm}, hereafter
called Ref. I, show that the isovector pairing correlations in the ground state of N=Z nuclei 
can be described with high precision by a condensate of alpha-like quartets built by 
collective  proton-neutron, 
neutron-neutron and proton-proton pairs. In this paper  we shall extend the calculation 
scheme of Ref. I  to nuclei away from the N=Z line and we will study to which extent 
alpha-like correlations coexist with the conventional pairing in nuclei with excess 
neutrons or protons. The possibility of coexistence/competition of 
four-particle correlations of alpha-type with the usual two-body pairing 
correlations was several times discussed
in the literature \cite{zelevinsky,eichler,dobes} but, as far as we know,
it was never checked in realistic microscopic calculations.  
 
In the present study  we consider a system of N neutrons and Z protons moving outside a 
self-conjugate core and interacting via  a charge-independent pairing force. 
The corresponding Hamiltonian is
\beq
\hat{H}= \sum_{i,\tau=\pm 1/2} \varepsilon_{i\tau} N_{i\tau}
+ \sum_{i,j,t=-1,0,1} V_{ij} P^+_{i,t} P_{j,t}
\eeq
where $\varepsilon_{i\tau}$ are the single-particle energies associated
to the mean fields of neutrons and protons, supposed invariant to time reversal.
The isovector interaction is expressed in terms of the isovector pair operators 
$P^+_{i,1}=\nu^+_i \nu^+_{\bar{i}}$, $P^+_{i,-1}=\pi^+_i \pi^+_{\bar{i}}$ and 
$P^+_{i,0}=(\nu^+_i \pi^+_{\bar{i}} + \pi^+_i \nu^+_{\bar{i}})/\sqrt{2}$; the
operators $\nu^+_i$ and $\pi^+_i$ create, respectively, a neutron and a proton in 
the state $i$ while $\bar{i}$ denotes the time conjugate of the state $i$.

In Ref. I the ground state of the hamiltonian (1) for a system with N=Z=even 
was  described by the trial state 
\begin{equation}
| \Psi \rangle =(A^+)^{n_q} |0 \rangle ,
\end{equation}
where $n_q=(N+Z)/4$ and $A^+$ is a collective four-nucleon operator defined by
\beq
A^+ = \sum_{i,j} x_{ij} A^+_{ij}.
\eeq
$A^+_{ij}$ denotes the non-collective four-nucleon operators  constructed 
by coupling two non-collective isovector pairs to the total isospin T=0, i.e.,
\beq
A^+_{ij} = [P^+_i P^+_j]^{T=0} = \frac{1}{\sqrt 3}
(P^+_{i,1} P^+_{j,-1}+P^+_{i,-1} P^+_{j,1}
           -P^+_{i,0} P^+_{j,0}).
\eeq
Supposing that the amplitudes $x_{ij}$ are separable,i.e., $x_{ij}=x_i x_j$,
the collective four-nucleon operator (3) can be written as
\beq
A^+= 2 \Gamma^+_1 \Gamma^+_{-1} - (\Gamma^+_0)^2,
\eeq
where $\Gamma^+_{t}= \sum_i x_i P^+_{i,t}$ denote, for t={0,1,-1},  the 
collective pair operators
for the proton-neutron (pn), neutron-neutron (nn) and proton-proton (pp)
pairs. Due to the isospin invariance, all the collective pairs have the same 
mixing amplitudes $x_i$. 

With the collective four-nucleon operator (5) the  state (2) can be written  as
\begin{eqnarray}
| \Psi \rangle &=& (2 \Gamma^+_1 \Gamma^+_{-1} - \Gamma^{+2}_0 )^{n_q} |0 \rangle \nonumber \\
         &=& \sum_k
\left(\begin{array}{l} n_q \\ k \end{array}\right)
(-1)^{n_q-k} 2^k (\Gamma^+_1 \Gamma^+_{-1})^k \Gamma^{+2(n_q-k)}_0 |0\rangle
\label{psi}
\end{eqnarray}
From the equation above it can be seen that the alpha-like condensate for a system 
with N=Z=even is a particular superposition of nn, pp and pn pair condensates.

Now we shall consider the case of even-even systems with an excess of one sort of
nucleons, neutrons or protons. For these systems we  suppose that
the excess neutrons or protons form a pair condensate of conventional type 
which is appended to  the alpha condensate. Thus, for even-even system with an excess 
of neutrons we consider the following ansatz for the ground state 
\begin{equation}
| \Psi \rangle = (\tilde{\Gamma}^+_1)^{n_N} (A^+)^{n_q} |0 \rangle = 
(\tilde{\Gamma}^+_1)^{n_N} (2 \Gamma^+_1 \Gamma^+_{-1} - \Gamma^{+2}_0 )^{n_q} |0 \rangle 
\end{equation}
where $n_N=(N-Z)/2$ is the number of neutron pairs in excess and $n_q=(N-2n_N+Z)/4$ is 
the maximum number of alpha-like quartets which can be formed by the neutrons and protons.
Since the quartets $A^+$ have zero isospin, the state (7) has a well-defined total
isospin given by the excess neutrons, i.e., T=$n_N$.  
The neutron pairs in excess are described by the collective pair operator 
$\tilde{\Gamma}^+_{1}= \sum_i y_i P^+_{i1}$. It can be seen that the collective
pair describing the excess neutrons is taken of different structure from the collective
neutron pair entering in the collective quartet. This is a requirement imposed by the 
Pauli principle in the HF limit. For the particular case of degenerate single-particle
states and a seniority-type pairing force the state (7) is the exact
solution of the Hamiltonian (1) \cite{dobes}.  

It is important to observe that in the state (7) one can identify two terms which play
the role of particle-number-projected BCS (PBCS) approximations for N$>$Z systems 
interacting with charge-independent pairing forces, i.e.,
\beq
|PBCS0 \rangle = (\tilde{\Gamma}^+_1)^{n_N} (\Gamma^+_0)^{2 n_q} |0 \rangle
\eeq
\beq
| PBCS1 \rangle = (\tilde{\Gamma}^+_1)^{N/2}  (\Gamma^+_{-1})^{Z/2} |0 \rangle .
\eeq
The state (8) is a product between a condensate of proton-neutron
pairs and a condensate of neutron-neutron pairs while the state (9) is a product 
of a condensate of neutron-neutron pairs with a condensate of proton-proton  
pairs. Both states have the right number of protons and neutrons but
have not a well-defined total isospin.

The mixing amplitudes $x_i$ and $y_i$ which define the ground state (7)
are determind from the minimization of $ \langle \Psi | H | \Psi \rangle$ under 
the normalization condition $ \langle \Psi | \Psi \rangle =1$. To calculate the
average of the hamiltonian and the norm we have extend the recurrence relations 
method of Ref. I by including the contribution of the excess neutrons. 
Thus the recurrence relations are calculated with the following states of arbitrary
numbers of collective nn, pp and np pairs
\beq
| n_1 n_2 n_3 n_4 \rangle = \Gamma^{+n_1}_1 \Gamma^{+n_2}_{-1} \Gamma^{+n_3}_0 
\tilde{\Gamma}^{+n_4}_1 |0 \rangle .
\eeq
Compared to N=Z systems, these states have two kind of neutron collective 
pairs, corresponding to the extra pairs and to the pairs which are included in the
quartet condensate. The recurrence relations satisfied by the matrix elements of 
the hamiltonian (1) with the states (10) can be simply related to the recurrence 
relations we have used in Ref. I for N=Z systems. 
Finally, we would like to stress that in the formalism presented here
the Pauli principle is incorporated rigorously, which is very important when
four-body correlations are calculated.

\begin{table}[hbt]
\caption{ Pairing correlations energies for isotopes having as core $^{16}$O.
 The results correspond to exact diagonalisation (Exact), quartet condensation 
model (QCM), and the PBCS1 approximation (9). 
Numbers in the brackets are the errors relative to the exact diagonalisation. 
The calculations are done with an isovector pairing force of seniority
type and with axially-deformed single-particle states.}
\begin{center}
\begin{tabular}{|c|c|c|c||c|c|c|c|}
\hline
\hline
   &    Exact & QCM & PBCS1 &  & Exact & QCM & PBCS1 \\
\hline
\hline
$^{20}$Ne  &  6.550    & 6.539  (0.17\%)  & 5.752 (12.18\%)  &
$^{24}$Mg  &  8.423   & 8.388 (0.41\%)  & 7.668 (8.96\%)  \\
$^{22}$Ne  &  6.997   &  6.969  (0.40\%)  & 6.600 (5.67\%)   &
$^{26}$Mg  &  8.680   & 8.654 (0.30\%)  & 8.258 (4.86\%)  \\
$^{24}$Ne  &  7.467   &  7.426  (0.55\%)  & 7.226 (3.23\%)   &
$^{28}$Mg  &  8.772   & 8.746 (0.30\%)  & 8.531 (2.75\%)  \\
$^{26}$Ne  &  7.626    & 7.592  (0.45\%)  & 7.486 (1.84\%)   &
$^{30}$Mg  &  8.672   & 8.656 (0.18\%)  & 8.551 (1.39\%)  \\
$^{28}$Ne  &  7.692    & 7.675  (0.22\%)  &  7.622 (0.91\%)  &
$^{32}$Mg  &  8.614   & 8.609 (0.06\%)  &  8.567 (0.55\%)   \\
$^{30}$Ne  &  7.997    & 7.994  (0.04\%)  &  7.973 (0.30\%)  &
$^{28}$Si  &  9.661   & 9.634 (0.28\%)  & 9.051 (6.31\%)  \\
$^{30}$Si  &  9.310   & 9.296 (0.15\%)  & 9.064 (2.64\%) & 
$^{32}$Si  &  9.292   & 9.283 (0.10\%)  & 9.196 (1.03\%)   \\
\hline
\hline
\end{tabular}
\end{center}
\end{table}

\begin{table}[hbt]
\caption{ The same as in Table I but for isotopes having as core $^{40}$Ca
and ${100}$Sn} 
\begin{center}
\begin{tabular}{|c|c|c|c|c|c|c|c|}
\hline
\hline
   &    Exact & QCM & PBCS1 &   &  Exact & QCM & PBCS1  \\
\hline
\hline
$^{44}$Ti  &  3.147   & 3.142 (0.16\%)  & 2.750 (12.61\%) & 
$^{48}$Cr  & 4.248   & 4.227 (0.49\%)   & 3.854 (9.27\%)   \\
$^{46}$Ti  &  3.526   & 3.509 (0.48\%)  & 3.308 (6.18\%)  &
$^{50}$Cr  & 4.461   & 4.444 (0.38\%)   & 4.230 (5.18\%)   \\
$^{48}$Ti  &  3.882   & 3.853 (0.75\%)  & 3.735 (3.79\%)  &
$^{52}$Cr  & 4.743   & 4.721 (0.46\%)   & 4.582 (3.39\%)   \\
$^{50}$Ti  &  3.973   & 3.956 (0.43\%)  & 3.889 (2.11\%)  &
$^{54}$Cr  &  4.869  & 4.855 (0.29\%)   & 4.772 (1.99\%)  \\
\hline
\hline
$^{104}$Te  &  1.084   & 1.082 (0.18\%) & 0.964 (11.07\%) &
$^{108}$Xe  &  1.870   & 1.863 (0.37\%) & 1.697 (9.25\%)  \\
$^{106}$Te  &  1.324   & 1.321 (0.23\%) & 1.250 (5.59\%)  &
$^{110}$Xe  &  2.191   & 2.185 (0.27\%) & 2.058 (6.07\%)  \\
$^{108}$Te  &  1.713   & 1.698 (0.88\%) & 1.642 (4.14\%)  &
$^{112}$Xe  &  2.449   & 2.437 (0.49\%) & 2.348 (4.12\%)  \\
$^{110}$Te  &  1.892   & 1.880 (0.63\%) & 1.843 (2.59\%) &
$^{114}$Xe  &  2.964   & 2.954 (0.34\%) & 2.887 (2.60\%)  \\
\hline
\hline
\end{tabular}
\end{center}
\end{table}

The model described above, which will be reffered to as quartet condensation 
model (QCM), as in Ref. I, is well-suited for studying the competition between the
alpha-like four-nucleon correlations and the conventional pairing condensation
in nuclei with proton-neutron pairing. As an illustration we apply it here  for 
three sets of nuclei with the valence nucleons moving outside the double-magic 
cores $^{16}$O, $^{40}$Ca and $^{100}$Sn, 
which are taken as inert. For each set of nuclei we start with the N=Z=even isotopes
and add extra neutron pairs. The calculations are done for those nuclei for which the 
ground state energy can be calculated exactly by  diagonalisation.
To check the accuracy of QCM we have done calculations using for
the single-particle energies and the pairing force the two different
inputs employed in Ref. I. Thus we first applied QCM  for a charge-independent
pairing interaction of seniority type, with the strength $g=24/A$, acting on 
protons and neutrons moving in deformed mean fields. The mean fields are 
obtained from axially-deformed Hartree-Fock (HF) calculations \cite{ev8} done with 
the Skyrme force SLy4 \cite{sly4}. 
From the HF spectrum of the three sets of nuclei we consider in the pairing 
calculations, respectively, the lowest 7, 9 and 10 states above the 
double-magic core. In the calculations we have neglected the 
Coulomb interaction and we have used for N$>$Z nuclei the same 
single-particle energies as for the corresponding N=Z isotope. As shown in 
Refs.\cite{neergard,bentley} , the isospin dependence of the single-particle 
energies can be eventually taken into account reasonably well by adding to the 
calculated binding energies a term proportional to T(T+1). 

 It is important to be mentioned that the Hamiltonian (1) 
with a similar input as employed here, using deformed mean fields provided by the 
Nilsson model instead of Skyrme-HF, is realistic enough for describing the experimental 
even-even to odd-odd energy difference as well as the term linear in 
N-Z (Wigner energy) in the nuclear binding energy \cite{bentley}.

The results we have obtained for pairing correlations energies with the input
presented above are shown in Tables I-II. The correlation energies are defined as 
$E_{corr}=E_0-E$, where $E$ is the total energy while $E_0$ is the energy 
obtained without the pairing interaction. In Tables I-II the QCM results are
compared to the exact results, obtained by direct diagonalisation, and with
the results provided by the PBCS1 approximation (9).  Since, as in N=Z systems
\cite{sandulescu_errea,sandulescu_qcm}, the PBCS0 approximation (8) gives less binding 
compared to PBCS1, its prediction are not given here.

Two points emerge immediately  from Tables I-II. First, it can be noticed 
that QCM describes with very good accuracy the pairing correlations 
energies for all calculated isotopes. Thus for all the isotopes the
errors  relative to the exact results (shown in the brackets) are
below $1\%$. Second, it can be seen that the PBCS1 
approximation, in which it is supposed that the system splits in two superfluid
composed by neutrons and protons, is  less accurate, much less than the PBCS
approximation for like-particle pairing \cite{sandulescu_bertsch}.
 As expected, its accuracy increases with the number of pairs in excess. 
Since by adding more and more neutron pairs the role of proton-neutron pairs 
is diminishing, one may think that there is a phase transition from a mixed 
condensate of alpha-like quartets and neutron pairs to a standard mixed 
condensate of neutrons and protons, as described by the state PBCS1. From the calculations
presented in Tables I-II it appears that this is not the case.

Apart from correlation energies, we have also checked that QCM predictes 
accurate results for occupation probabilities of single-particle states. 
As an example in Table III are shown the results for the isotope $^{30}$Mg. 

\begin{table}[hbt]
\caption{ Occupation probabilities of  single-particle states in  $^{30}$Mg.
Are shown the exact and the QCM results for neutrons (n) and protons (p).  }
\begin{center}
\begin{tabular}{|c|c|c|c|c|}
\hline
\hline
      $\varepsilon_i$  &  Exact(n)   &   QCM(n)   &   Exact(p)  &  QCM(p)  \\
\hline
\hline
    -16.45   &  0.995   & 0.995  &  0.983    & 0.983   \\
    -13.94   &  0.993   & 0.993  &  0.961    & 0.963   \\
    -10.39   &  0.987   & 0.987  &  0.028    & 0.026   \\
     -8.08   &  0.971   & 0.972  &  0.012    & 0.017   \\
     -6.09   &  0.921   & 0.923  &  0.007    & 0.007   \\
     -3.89   &  0.087   & 0.085  &  0.005    & 0.005   \\
     -2.61   &  0.045   & 0.045  &  0.004    & 0.004   \\
\hline
\hline
\end{tabular}
\end{center}
\end{table}

More specific informations about the correlations described by QCM
can be extracted from the entanglement properties of the Cooper pairs
which compose the ground states (7). As a measure of the entanglement 
we use here the  so-called Schmidt number \cite{schmidt} defined as
$K=(\sum_i w_i^2)^2/\sum_i w_i^4$ , where $w_i$ are the mixing amplitudes of the
two-body function which describes the entangled particles  
(for an application of Schmidt number to like-particle pairing in nuclei 
see \cite{sandulescu_bertsch}). In the case of the Cooper pairs $\Gamma^+_t$ and 
$\tilde{\Gamma}^+_1$ the mixing amplitudes $w_i$ are, respectively, $x_i$ and $y_i$. 
As expected, the Schmidt numbers show that the entanglement of the  proton 
pairs is stronger when they are included in the quartets than when they 
form a  pair condensate as in PBCS1. For example, in $^{30}$Mg we obtain K= 1.88 for 
the protons in the quartet condensate and K=1.79 for the protons in the pair condensate. 
As for the neutron pairs in excess, they are usually much more entangled than the 
ones included in the quartets (e.g., by about 64$\%$  in  $^{30}$Mg).

To check further the accuracy of QCM, we have also done  
calculations with more general isovector pairing forces, extracted from 
the (T=1,J=0) part of standard shell model interactions, acting on 
spherical single-particle states. As an example, in Table IV we present 
the correlation pairing energies obtained for the  nuclei having as 
closed core $^{100}$Sn. One can observe that QCM gives very good predictions, 
comparable to the calculations done with the seniority-type force presented 
in Tables I-II. The calculations have been  done with the isovector pairing 
force extracted from the effective G-matrix interaction of Ref. \cite{bonnA} 
and with the single-particle energies $\varepsilon_{2d_{5/2}}$=0.0, 
$\varepsilon_{1g_{7/2}}$=0.2,
$\varepsilon_{2d_{3/2}}$=1.5, $\varepsilon_{3s_{1/2}}$=2.8. The intruder state
$h_{11/2}$ was not introduced in the calculations because with it 
the exact diagonalisations cannot be performed due to the very large matrices  
(e.g,  186 billions for $^{116}$Xe). It is worth mentioning that the intruder
state, which has a significant influence for the heavier isotopes shown in Table IV, 
can be simply accounted for in QCM.  In fact, as any approach based on variational
principle, the QCM can be applied  for  nuclei and model spaces which are far beyond 
the capability of present shell model codes.

\begin{table}[hbt]
\caption{ Pairing correlations energies for isotopes having as core 
$^{100}$Sn calculated with the isovector pairing force extracted from the 
effective G-matrix interaction of Ref. \cite{bonnA} and with spherical 
single-particle states. The notations are the same as in Table I.}
\begin{center}
\begin{tabular}{|c|c|c|c|c|c|c|c|}
\hline
\hline
   &    Exact & QCM & PBCS1 &  & Exact & QCM & PBCS1 \\
\hline
\hline
$^{104}$Te &  3.831    & 3.829 (0.05\%) & 3.607 (5.85\%)  &
$^{108}$Xe &  6.752    & 6.696 (0.83\%) & 6.311 (6.53\%)  \\
$^{106}$Te &  5.156    & 5.130 (0.50\%) & 4.937 (4.25\%)  &
$^{110}$Xe &  7.578    & 7.509 (0.91\%) & 7.184 (5.20\%)  \\
$^{108}$Te &  5.970    & 5.930 (0.67\%) & 5.768 (3.38\%)  &
$^{112}$Xe &  8.285    & 8.208 (0.93\%) & 7.944 (4.12\%)  \\
$^{110}$Te &  6.664    & 6.616 (0.72\%) & 6.485 (2.69\%)  &
$^{114}$Xe &  8.446    & 8.368 (0.92\%) & 8.167 (3.30\%)  \\
$^{112}$Te &  6.815    & 6.764 (0.75\%)  & 6.665 (2.20\%)  &
$^{116}$Xe &  8.031    & 7.947 (1.05\%)  & 7.810 (2.75\%)  \\
\hline
\hline
\hline
\end{tabular}
\end{center}
\end{table}

To conclude, in this paper we have shown that four-nucleon correlations of 
alpha-type are very important in systems with neutron-proton pairing. 
This is true not only for systems with N=Z but also for systems with 
excess neutrons. It means that, whenever possible,  the protons and
neutrons prefer to couple together in alpha-like quartets which are forming
an alpha-like condensate. When not all neutrons can be included in the
alpha-like quartets, the excess neutrons form a typical condensate of 
collective pairs which is appended to the alpha-like condensate. 
We have found that the alpha-type correlations coexist with the conventional 
pairing of excess neutrons irrespective to the number of excess neutrons.  
To the best of our knowledge, these are the first realistic microscopic 
calculations which point to the coexistence of alpha-like quartets and 
conventional Cooper pairs in nuclei away of N=Z line.
 
\vskip 0.2cm
\noindent
{\bf Acknowledgements}
\vskip 0.1cm
\noindent

N.S. thanks the hospitality of Royal Institute of Technology, Stockholm,
where this article has been  written. This work was supported by the 
Romanian Ministry of Education and Research through the grant Idei nr 57, 
by the U.S. Department of Energy through the grant DE-FG02-96ER40985 and by 
IN2P3-NIPNE collaboration . D. N. acknowledges the support from the PhD grant 
POSDRU/107/1.5/S/82514.

\end{document}